\documentclass[aps,prd,10pt,nofootinbib,twocolumn,eqsecnum,showpacs,showkeys,superscriptaddress,preprintnumbers,altaffilletter]{revtex4-1}

\usepackage{graphicx}
\usepackage{amsmath}
\usepackage{multirow}
\usepackage{hyperref}
\usepackage{soul}
\usepackage{graphicx}
\usepackage{dcolumn}
\usepackage{amssymb}
\usepackage{amsfonts}
\usepackage{amsbsy}
\usepackage{color}
\usepackage{rotating}
\usepackage[english]{babel}
\usepackage{multirow}
\usepackage{float}

\newcommand{\be}{\begin{equation}}
\newcommand{\ee}{\end{equation}}
\newcommand{\bea}{\begin{eqnarray}}
\newcommand{\eea}{\end{eqnarray}}

\hypersetup{
    colorlinks=true,
    linkcolor=blue,
    filecolor=magenta,
    urlcolor=cyan,
    citecolor=cyan
}

\begin{document}

\title{On the foundations of entropic cosmologies: inconsistencies, possible solutions and dead end signs.}
\author{Hussain Gohar}
\email{hussain.gohar@usz.edu.pl}
\affiliation{Institute of Physics, University of Szczecin, Wielkopolska 15, 70-451 Szczecin, Poland}
\author{Vincenzo Salzano}
\email{vincenzo.salzano@usz.edu.pl}
\affiliation{Institute of Physics, University of Szczecin, Wielkopolska 15, 70-451 Szczecin, Poland}

\date{\today}

\begin{abstract}
In this letter we explore the foundations of entropic cosmology and highlight some important flaws which have emerged and adopted in the recent literature.
We argue that, when applying entropy and temperature on the cosmological horizon by assuming the holographic principle for all thermodynamic approaches to cosmology and gravity, one must derive the consistent thermodynamic quantities following Clausius relation. One key assumption which is generally overlooked, is that in this process one must assume a mass-to-horizon relation, which is generally taken as a linear one. We show that, regardless of the type of entropy chosen on the cosmological horizon, when a thermodynamically consistent corresponding temperature is considered, all modified entropic force models are equivalent to and indistinguishable from the original entropic force models based on standard Bekenstein entropy and Hawking temperature. As such, they are also plagued by the same problems and inability to describe in a satisfactory qualitative and quantitative way the cosmological dynamics as it emerges from the probes we have. We also show that the standard accepted parameterization for Hawking temperature (including a $\gamma$ rescaling) is actually not correctly applied, namely, it is not related to entropy in a thermodynamically consistent way. Finally, we clearly state that the explicit form of the entropic force on cosmological horizons is mostly dictated by the assumption on the mass-to-horizon relation. As such, we discuss what should be done in order to fix all such issues, and what conceptually could be implied by its correct implementation in order to advance in the field.
\end{abstract}

\maketitle

\section{\label{sec:level1}Introduction}


Current observational data reveal that the Universe's expansion is accelerating \cite{Perlmutter:1998np,Riess:1998cb,Hinshaw:2012aka,Aghanim:2018eyx,eBOSS:2020yzd,Brout:2022vxf,DES:2022ccp}, and numerous investigations have been conducted to better comprehend this observational fact. The most conventional way for explaining the accelerated expansion of the Universe involves the addition and inclusion of a new element, the so-called dark energy fluid \cite{Li:2011sd}, to the cosmic inventory. This fluid has peculiar features in comparison to other standard energy-matter contributions, and its effects on the Einstein field equations are precisely set to cause such accelerated expansion. 

Despite their many accomplishments, dark energy models are unable to adequately explain all of the observational evidence that we have gathered so far and are also impacted by several additional theoretical flaws \cite{Bull:2015stt}. As a result, numerous different theoretical ideas have been proposed in order to address and comprehend its nature \cite{Nojiri:2006ri,Clifton:2011jh,Ishak:2018his}. 

In this work, we will pay special attention to the seminal papers of Easson et al. \cite{Easson:2010av,Easson:2010xf}, who offered entropic cosmology as a fresh explanation for the accelerated expansion. The entropic cosmology approach requires the inclusion of additional entropic force terms to the Einstein field equations, assumed to come from the boundary terms in the Einstein Hilbert action \cite{Easson:2010av, Easson:2010xf,Espinosa-Portales:2021cac,Garcia-Bellido:2021idr}. 

The physical rationale for the entropic force stems from the holographic principle, which states that the degrees of freedom in the bulk of the Universe can be stored at/onto the Universe's boundary \cite{tHooft:1993dmi, Susskind:1994vu}. Hence, by associating an entropy and a temperature on/to the horizon, one can define an entropic force acting on the boundary of the Universe, which might be responsible for the Universe dynamics on large cosmological scales and eventually provide a solution for its accelerated expansion \cite{Komatsu:2012zh,Komatsu:2013qia,Komatsu:2014lsa,Komatsu:2014vna,Komatsu:2014ywa,Komatsu:2015nga,Komatsu:2015nkb,Gohar:2020bod,Koivisto:2010tb,Cai:2010kp,Cai:2010zw}. 

Bekenstein \cite{Bekenstein:1973ur} and Hawking \cite{Hawking:1974rv} demonstrated in the 1970s that the principles of black hole thermodynamics are similar to the laws of ordinary thermodynamics . Since then, a wide variety of gravitational and cosmological applications have been explored by utilizing the relationship between the area of the event horizon and the surface gravity of the black hole with the entropy and temperature in conventional thermodynamics. Jacobson \cite{Jacobson:1995ab} obtained the Einstein field equations from the entropy and horizon area proportionality by assuming the flow of heat across the horizon, and the Friedmann and acceleration equations were derived by Padmanabhan \cite{Padmanabhan:2003pk, Padmanabhan:2009vy} using the holographic equipartition law, which states that the difference between the degrees of freedom at the surface and in the bulk of an area of space is what causes the expansion of cosmic space. Furthermore, Verlinde \cite{Verlinde:2010hp} proposed a new concept: gravity is defined as an entropic force that arises in a system as a result of the statistical tendency to increase its entropy.  All of these novel perspectives offer fresh insights into the quantum gravity conundrum and may even examine the thermodynamic origins of space-time. All of these approaches rely on the Bekenstein entropy and Hawking temperature, which satisfy the Clausius relation under the presumption that mass scales with the horizon linearly.

In entropic cosmology, the form of the driving entropic force terms is determined by the definition of the entropy. In the original entropic force models, the Bekenstein entropy \cite{Bekenstein:1973ur} together with the Hawking temperature \cite{Hawking:1974rv} has been used to get the entropic force terms. However, such models fail to account for both the Universe's acceleration and deceleration and it has been shown in \cite{Basilakos:2012ra,Basilakos:2014tha} that they neither account for cosmic fluctuations nor are compatible with the formation of structures.

In recent years, nonextensive thermodynamic applications have been used in gravitation and cosmology \cite{Czinner:2015ena,Czinner:2015eyk,Czinner:2017bwc,Czinner:2017tjq,Alonso-Serrano:2020hpb,Nojiri:2021czz,Nojiri:2021jxf,Nojiri:2022aof,Nojiri:2022ljp,Nojiri:2022sfd,Nojiri:2021czz,Nojiri:2022aof,Nojiri:2022sfd,Nojiri:2021jxf,Cimdiker:2022ics,Cimidiker:2023kle,Promsiri:2020jga,Promsiri:2021hhv,Nakarachinda:2021jxd,Promsiri:2022qin,Nakarachinda:2022gsb,Saridakis:2020zol,Dabrowski:2020atl,Nojiri:2022ljp,Komatsu:2016vof,Komatsu:2015nkb,Nunes:2015xsa,Liu:2022snq,Majhi:2017zao,Luciano:2021mto,DiGennaro:2022ykp,DiGennaro:2022grw,Asghari:2021bqa,Abreu:2022pil,Komatsu:2016vof}. The use of nonextensive entropies for self-gravitating systems is obvious given the crucial role that long-range interactions play in nonextensive thermodynamics. There are several different ``concepts'' of nonextensive entropies that have been proposed in this regard, including Tsallis \cite{Tsallis:1987eu,tsallisbook}, R\'enyi \cite{reny1}, Sharma-Mittal \cite{SM2,SM,SM1}, Kaniadakis \cite{KD}, Tsallis-Cirto \cite{Tsallis:2012js,Tsallis:2019giw}, Tsallis-Zamora \cite{Zamora:2022cqz,Zamora:2022sya}, and Barrow \cite{Barrow:2020tzx} entropies.

Tsallis entropy generalizes Gibbs/Shannon entropy for scenarios in which additive and extensive property of entropy do not work. From the standpoint of quantum information, R\'enyi entropy measures the entanglement of quantum systems and has thus been applied to black holes and cosmological horizons. Similarly Tsallis-Cirto entropy is motivated to make the black hole entropy extensive. Meanwhile, Barrow entropy is mathematically equivalent to Tsallis-Cirto entropy but is entirely motivated by the fractal structure of the horizon due to quantum fluctuations. In \cite{Komatsu:2012zh,Komatsu:2013qia}, the authors adapted a particular case of Tsallis-Cirto entropy and a higher dimensional entropy, known as quartic entropy, to the cosmological horizons within the context of entropic cosmology and addressed the problem of a decelerating and accelerating Universe, as well as the issue of structure formation. In addition, Sharma-Mittal entropy generalizes the Tsallis and R\'enyi entropies, whereas Kaniadakis entropy is inspired by special relativity considerations.

In the context of black holes and cosmology, defining different nonextensive entropies on black hole and cosmological horizons and their associated consistent thermodynamic quantities is crucial from a thermodynamic perspective \cite{Cimdiker:2022ics,Nojiri:2021czz}. Let's use the Bekenstein entropy and Hawking temperature as an example. If we apply Clausius's relation to this scenario, it provides a consistent thermodynamic mass for a Schwarzschild black hole that demonstrates the equality of mass and energy. Clausius's relation, on the other hand, gives an inconsistent mass-energy relationship when using Hawking temperature with R\'enyi entropy or Tsallis-Cirto entropy \cite{Nojiri:2021czz}.
Nonetheless, Clausius's relation provides consistent mass-energy relations for both scenarios using the consistent temperatures corresponding to R\'enyi and Tsallis-Cirto entropies \cite{Cimdiker:2022ics}. The Bekenstein entropy has a challenge in that we only employ the entropy-area relation, which is supported by the Hawking's area theorem \cite{Hawking:1971tu, Hawking:1971vc}, and we need a statistical mechanics description of black hole entropy to understand the black hole microstates \cite{Strominger:1996sh}. Quantum field theory, however, validates the Hawking evaporation process for determining the Hawking temperature, which further supports the idea of black hole entropy as the area of the black hole's horizon.

The paper is organized as follows. In Sec.~(\ref{sec:entropic_review}) we will provide a brief summary of the main way to proceed in order to construct a proper entropic cosmology model. We will highlight how some parameters which are commonly defined in literature are actually misused leading to flaws in the same foundation of entropic cosmologies. In Sec.~(\ref{sec:nonextensive}) we introduce several thermodynamic approaches in the nonextensive setup which have been recently used to investigate the evolution of the Universe. We will discuss how these approaches are fundamentally wrong and how one should properly face them. Finally, in Sec.~(\ref{sec:discussion}) we will expose our final summary and conclusions.

\section{Entropic Cosmology Review: the Bekenstein-Hawking standard}
\label{sec:entropic_review}

In this work, we take into account a flat, homogeneous and isotropic universe. The Bekenstein entropy and Hawking temperature on the horizon $L$, indicated respectively as $S_{BH}$ and $T_{BH}$, assuming the holographic principle, read as 
\begin{align}
S_{BH}&=\frac{k_Bc^3\, A}{4\hbar G}, \label{entropy1} \\
T_{BH}&=\frac{\hbar c }{2\pi k_B L\, } \label{temperature1}, 
\end{align}
where: $G$, $c$, $\hbar$ and $k_B$ are the Newton's gravitational constant, the speed of light, the reduced Planck's constant, and the Boltzmann's constant, respectively; the surface gravity is $\kappa = 1/L$ on the cosmological horizon;  $A$ is the surface area
\begin{equation}\label{area}
A=4\pi L^2
\end{equation}
whose radius is the cosmological horizon $L$, which is \textit{generally} assumed to be the Hubble horizon,
\begin{equation}\label{horizon}
L=\frac{c}{H}\, ,
\end{equation}
with the Hubble parameter defined as $H=\dot a/a$, where $a=a(t)$ is the scale factor at time $t$ and a dot denotes derivative with respect to time $t$.

It is worth noting that the Hubble horizon $L$ equals the apparent horizon $r_A = 1/\sqrt{H^2+k/a^2}$ for a flat universe $(k=0)$, but in general for an accelerated expanding universe we can also have an event horizon $r_E = a(t)\int_t^\infty \frac{dt}{a(t)}$ \cite{Faraoni:2011hf,Davis:2003ad}. Only in a flat de Sitter universe the Hubble, apparent, and cosmological event horizons coincide and all have the same constant value of $1/H$. In general, in an expanding universe (with positive dark energy density), the Hubble horizon only asymptotically coincide with the event horizon. The issue with $r_E$ is that it does not always exist in FLRW universes, whereas the apparent and Hubble horizons do. Furthermore, for dynamical spacetime, it has been demonstrated that the apparent horizon is the causal horizon, with an associated gravitational entropy and surface gravity for the apparent horizon \cite{Cai:2008gw,Li:2008gf}.
Given that the universe is spatially flat, isotropic, and homogeneous, it makes sense to regard the apparent horizon — in this case, the Hubble horizon — as the universe's boundary or horizon. It is demonstrated that in a four-dimensional de Sitter space \cite{Gibbons:1977mu}, the horizon area $A = 4 \pi/ H^2$ is related to the entropy, and that an observer at the origin detects a thermal radiation from the de Sitter horizon at $R=1/H $ with the temperature $T = H/2\pi$.

Furthermore, for the dynamic apparent horizon, it is clear that in order to calculate surface gravity, one must know not only the apparent horizon radius and the Hubble parameter, but also the time-dependence of the horizon radius. As a result, when applying the first law to the apparent horizon to calculate surface gravity and thus temperature, and considering an infinitesimal amount of energy crossing the apparent horizon, the apparent horizon radius $r_A$ should be regarded as having a fixed value \cite{Cai:2005ra,Bousso:2004tv,Viaggiu:2015cra}. Thus, the relation $T\approx|\kappa|/(2\pi)=1/(2 \pi r_A)$ between the temperature and surface gravity at the apparent horizon is recovered, and we have $\kappa \approx -1/r_A$. In this way, we can justify the temperature assumption in Eq.~(\ref{temperature1}). In other words, the first law of thermodynamics may only hold approximately for the dynamical apparent horizon. 

Although, it seems there is no final consensus on the issue. Indeed, in \cite{Bousso:2004tv} it is argued that a thermodynamic description of the horizon is approximately valid, so it doesn't really matter whether one uses the apparent horizon or the event horizon. But in \cite{Viaggiu:2015cra}, it is shown that the temperature defined at the apparent horizon is the consequence of the ordinary formulas of thermodynamics and, therefore, not the outcome of an approximation. While in \cite{Wang:2005pk} it is stated that the cosmological event horizon (in a universe filled with a dark energy fluid) might be unphysical from the point of view of the laws of thermodynamics. 

It is also widely accepted that the black hole entropy area formula is only valid in Einstein's General Relativity, that is, when the action of gravity theory contains only a linear term of scalar curvature and Hawking temperature in terms of surface gravity is independent of any theory. Bekenstein entropy has been modified for many gravity modifications and applied to many cosmological applications \cite{Cai:2005ra}. 

The intriguing query that arises when we apply definitions of entropies to black hole and cosmological horizons from various perspectives in thermodynamics and statistical mechanics is: what principles should be followed in order to define these entropies and the corresponding thermodynamic quantities on the horizons, given that we are unaware of any background geometrical theory related to these definitions of entropies? Therefore,
while applying to cosmological and black hole horizons, entropy and temperature are connected in at least two important ways. First of all, in order to prevent thermodynamic inconsistencies, the two quantities \textit{must} adhere to the Clausius relation,
\begin{equation}
dE = c^{2}\, dM = T\, dS\,  \label{eq:clausius}
\end{equation}
where we have used the mass-energy relation $E = Mc^2$.

Secondly, what is also \textit{fundamentally crucial} in and for all the further steps is the mass-horizon $(M-L)$ relation, which is \textit{unknown} a priori. The common procedure is to assume, \textit{by analogy with black holes}, that the mass scales linearly with the cosmological horizon as
\begin{equation}
M = \frac{c^2}{G} L\, . \label{eq:mass}
\end{equation}
This relation is crucial for defining temperature and entropy on the cosmological horizon which are at the same time satisfying the Clausius relation. 
It can be easily verified that the pair $\{S_{BH},T_{BH}\}$ from Eqs.~(\ref{entropy1})~-~(\ref{temperature1}) is consistent with Eq.~(\ref{eq:clausius}) when adopting the relation Eq.~(\ref{eq:mass}).
When we say consistent, we mean that the area law for the Bekenstein-Hawking entropy can always be obtained if we identify the thermodynamical energy $E$ with the black hole mass $M$ and the system temperature with the Hawking temperature, $T = T_{BH}$. Thus, using the Clausius relation, Eq.~ (\ref{eq:clausius}), with the $(M-L)$  relation, Eq.~(\ref{eq:mass}), yields $S_{BH}$. This also implies that the mass-energy relation associated with the horizon should be given by the Clausius relation with mass-horizon relation when temperature and entropy are defined on the horizon using the holographic principle. The Bekenstein-Hawking case makes this clear, but the Clausius relation results in an inconsistent mass-energy relation for the horizons when Hawking temperature is introduced along with nonextensive entropies on the horizons \cite{Nojiri:2021czz,Cimdiker:2022ics}. This means that, aside from Bekenstein entropy, the Hawking temperature is inconsistent with nonextensive entropies \cite{Cimdiker:2022ics}. 

Naturally, the question now emerges: can we, on the basis of the holographic principle, use nonextensive entropies on the horizons other than Bekenstein entropy? There are two ways we can respond to this query. 

First, we must rely on the thermodynamic definition of nonextensive entropies and, using the Clausius and mass horizon relations, obtain the associated temperatures (which will differ from the Hawking temperature) \cite{Cimidiker:2023kle}. We can avoid inconsistencies this way. However, providing physical justifications for these temperatures is difficult because they cannot be justified by quantum field theory. 

Second, because the Hawking temperature is the only justified temperature on the horizon in terms of surface gravity, one must be careful to define the consistent thermodynamic quantities while using Hawking temperature on the horizon within the nonextensive setup in order to obtain the correct mass-energy relation that is consistent with the holographic principle. In order to tackle this issue, recently, we have defined a generalized horizon entropy, which yields consistent thermodynamic quantities with Hawking temperature within nonextensive setup, based on a generalized mass to horizon relation that we recently provided in \cite{HG}.

The mass-horizon relation Eq.~(\ref{eq:mass}) is, in fact, a crucial premise to apply consistent thermodynamic quantities on the cosmological horizons from a thermodynamic perspective. Geometrically speaking, Eq.~(\ref{eq:mass})- with a factor $1/2$ - is the definition of Misner-Sharp mass/energy for a spherically symmetric case \cite{Gong:2007md}, defined for marginal-trapped horizon and the apparent horizon is an example of a trapped null surface \cite{Hayward:1994bu}. This validates the mass-horizon relation assumption needed to use the thermodynamic quantities in line with the holographic principle. 

Nevertheless, our aim in this paper is not to provide geometrical descriptions of these quantities, but rather more thermodynamic ones. Interestingly, in \cite{Gong:2007md}, the geometrical definition of first law or Clausius relation at the apparent horizon has been investigated for different theories of gravity from a mass-like function (see for example Eq.~(11) of \cite{Gong:2007md}), which reduces to be Misner-Sharp mass/energy for a spherically symmetric spacetime at the apparent horizon. But, here, we considered the Clausius relation in more conventional thermodynamic sense, and Eq.~(\ref{eq:mass}) as an assumption to be applied in/with the Clausius relation.

With these premises, one then proceeds to define, on the cosmological horizon, the entropic force $F_{BH}$ as
\be
F_{BH}=-T_{BH}\frac{dS_{BH}}{dL}, \label{entropicforce}
\ee
where the direction of increasing entropy is indicated by the minus sign. 
Combining Eqs.~(\ref{entropy1})~-~(\ref{temperature1})~-~(\ref{area})~-~(\ref{horizon}) with Eq.~(\ref{entropicforce}), we find that  the entropic force on the Hubble horizon is
\be
F_{BH}=-\frac{c^4}{G}\,. \label{entropicforce1}
\ee
Within entropic cosmology approaches, this force is believed to be responsible for the Universe's accelerating expansion. It is worth noting that the entropic force $F_{BH}$ can be related with the maximum force, $F_{max}= c^4/4G$ in general relativity by multiplying $F_{BH}$ with a factor of 1/4 \cite{Schiller:2005eme,Gibbons:2002iv,Barrow:2014cga,Dabrowski:2015eea,Ong:2018xna,Schiller:2021jxf,Jowsey:2021ixg}.
Likewise, the entropic pressure $p_{e}$ on the Hubble horizon caused by the entropic force $F_{BH}$ can be expressed as
\be
p_{e}=\frac{F_{BH}}{A}=-\frac{c^2}{4\pi G}H^2. \label{entropicpressure}
\ee

Now the first important point to highlight is that in literature the Hawking temperature on cosmological horizons (more specifically, on Hubble horizon) is more generally used, together with Bekenstein entropy $S_{BH}$ from Eq.~(\ref{entropy1}), written as
\begin{equation}\label{temperature_gamma}
T^{\gamma}_{BH}=\gamma\frac{\hbar c}{2\pi k_B L},
\end{equation}
where $\gamma$ is a non-negative dimensionless free parameter that theoretical considerations assume to be of the order of one \cite{Easson:2010av,Easson:2010xf} and whose nature and need is usually not clearly stated. 

In the original proposals, \cite{CaiSaridakis:2010kp,Qiu:2011zr}, later echoed by \cite{Komatsu:2012zh,Komatsu:2013qia,Komatsu:2014vna}, the $\gamma$ is used to \textit{quantify our ignorance about the exact location of the cosmological horizon}. Indeed, this is generally assumed to be the Hubble horizon, as in Eq.~(\ref{horizon}), but this is actually just an assumption. Thus, $\gamma$ should express how close is the \textit{true (unknown)} horizon to the Hubble one. But there are at least three important points which should be discussed about these definitions. 

The first one relates to the nature of the Hubble horizon as a \textit{true horizon}. If by horizon we mean a border which cannot be crossed and can be approached only for $t\rightarrow \infty$, then, the Hubble horizon is \textit{not a true horizon}, as well described in \cite{Davis:2003ad}. 

This consideration directly leads to the second point to be clarified: if a true horizon, like the event or particle horizon, should be considered, than the constant parameter $\gamma$ is unable to describe the variety of possible horizons when applied (and defined) with respect to the Hubble horizon. Namely, \textit{we cannot move} from the Hubble horizon to the event or particle horizon just by a constant factor $\gamma$, being each and every horizon defined in a different way with respect to each other (see again \cite{Davis:2003ad} for a quick compendium).

Finally, let us discuss the third point which is even more important in the context of entropic cosmologies and cosmological horizons, is independent of the definition of the latters one, and is generally overlooked: \textit{if} with the parameter $\gamma$ we are basically ``rescaling'' the cosmological horizon where the Hawking temperature is defined, as $L \rightarrow L/\gamma$, this \textit{should also influence} the expression of the Bekenstein entropy, Eq.~(\ref{entropy1}), which \textit{must be calculated on the same surface area}, and of the mass, \textit{if it scales with the horizon length} as in Eq.~(\ref{eq:mass}).

Indeed, one can easily check that using $T^{\gamma}_{BH}= \gamma\, T_{BH}$, as from Eq.~(\ref{temperature_gamma}), with $S_{BH}$ leads to a mass
\begin{align}
M&= \gamma \frac{c^2}{G}\, L,
\end{align}
which is clearly \textit{not consistent} with the hypothesis that $M\propto L/\gamma$. Instead, using the correct pair of thermodynamical quantities, $T^{\gamma}_{BH}$, and $S^{\gamma}_{BH}$ defined as
\begin{align}\label{area_gamma}
S^{\gamma}_{BH}&= \frac{S_{BH}}{\gamma^2}\, ,
\end{align}
we finally have the consistent mass:
\begin{align}\label{mass_gamma}
M&= \frac{1}{\gamma}\frac{c^2}{G}L.
\end{align}

If we calculate the entropic force as in Eq.~(\ref{entropicforce1}), we get
\be
F^{\gamma}_{BH}=-T^{\gamma}_{BH}\frac{dS^{\gamma}_{BH}}{dL}=-\frac{c^4}{G}\,, \label{entropicforce_gamma}
\ee
namely, $F^{\gamma}_{BH}=F_{BH} $. \textit{But} the entropic pressure is going to change, because
\be
p^{\gamma}_{e}=\frac{F^{\gamma}_{BH}}{A}=-\gamma^2 \frac{c^2}{4\pi G}H^2. \label{entropicpressure_gamma}
\ee

Furthermore, one might ask if the standard choice made in literature of working with the pair $\{S_{BH},T^{\gamma}_{BH}\}$ is \textit{definitely} wrong, as we have just shown. The answer is that, theoretically, they could still be consistent, \textit{if and only if we modify} Clausius equation. Indeed, Eq.~(\ref{eq:mass}) is \textit{just an assumption} and, we may add, \textit{not the most general one}. In fact, we might easily generalize it to
\begin{equation}
M_{\gamma} = \gamma \frac{c^2}{G} L\, , \label{eq:mass_gamma}
\end{equation}
with $\gamma$ controlling the scaling between the mass and the cosmological horizon. With $\gamma=1/2$ one would recover the standard black hole formulation (in geometrical sense, Misner-Sharp mass for spherically symmetric case), while its value would be unknown at cosmological scales and to be fixed by theoretical considerations and by comparison with observational data. Moreover, with Eq.~(\ref{eq:mass_gamma}), it can be easily checked that $S_{BH}$ and $T^{\gamma}_{BH}$ make a consistent thermodynamical set of physical quantities, which for clarity we will indicate as
\begin{align}
S^{M_\gamma}_{BH}&= S_{BH} \label{S_Bek_Mgamma}\\
T^{M_\gamma}_{BH}&=\gamma\, T_{BH}\, , \label{T_Bek_Mgamma}\,
\end{align}
but now with a \textit{quite clear difference} in the physical meaning of the dimensionless constant $\gamma$, which does not reflect anymore uncertainty in the horizon, but uncertainty in the $M-L$ relation. This interpretation is also justified by the argument that the thermodynamics of apparent horizons are approximately defined \cite{Bousso:2004tv}. In this case, the entropic force becomes
\be
F^{M_\gamma}_{BH}=-T^{\gamma}_{BH}\frac{dS_{BH}}{dL} = -\gamma \frac{c^4}{G}\,, \label{entropicforce_Mgamma}
\ee
and the corresponding entropic pressure $p^{M_\gamma}_{e}$ on the Hubble horizon is
\be
p^{M_\gamma}_{e}=\frac{F^{M_\gamma}_{BH}}{A}=-\gamma\frac{c^2}{4\pi G}H^2\, .\label{entropicpressure_Mgamma}
\ee
Just for the sake of clarity, here and in the next sections, where more entropy definitions ($S_i$) will be introduced, we name and refer to each of the previous scenarios as: \textit{standard}, when using $\{S_i,T_i\}$; \textit{horizon-scaled}, when using $\{S^{\gamma}_i,T^{\gamma}_i\}$; and \textit{mass-scaled} when using $\{S^{M_\gamma}_i,T^{M_\gamma}_i\}$.

\section{Entropic Cosmology Updated: nonextensive entropies}
\label{sec:nonextensive}

With the development of nonextensive thermodynamics and its applications to cosmology and gravitation, a large number of investigations have been done using different forms of nonextensive entropies on black hole and cosmological horizons. Particularly, on cosmological horizons, Hawking temperature has been used with these different nonextensive entropy forms, which is, from a thermodynamic point of view, inconsistent with nonextensive entropies other than Bekenstein entropy. For instance, in the setting of entropic cosmology, the authors in \cite{Komatsu:2013qia} combined $T_{BH}$ with a particular case of Tsallis-Cirto entropy to obtain entropic force $F_{TC} \propto L$, which shows that entropic force is directly proportional to the horizon $L = c/H$. Hence, they have different entropic force models with these inconsistent assumptions. Similarly, in \cite{Sheykhi:2018dpn}, the modified Friedmann equation was obtained using the first law of thermodynamics by defining the Hawking temperature $T_{BH}$ on the apparent horizon and connecting it to the Tsallis-Cirto entropy $S_{TC}$. In this way, they investigated the emergence of spacetime by using the holographic equipartition law, introduced by Padmanabhan, with Tsallis-Cirto entropy and Hawking temperature. The same approach is used in \cite{Komatsu:2016vof} by using R\'enyi entropy with Hawking temperature to get a cosmological constant-like term in the Friedmann equations.

There are many considerations like these in the literature and the main aim and goal of this section is to clearly state a point which might sound obvious but it is systematically disregarded and which we anticipate here: one should always use \textit{only the proper and consistently defined (i.e by Clausius relation) pairs of thermodynamic quantities} in any consideration. Moreover, we anticipate here that \textit{when this is done, all of the nonextensive entropic cosmologies we consider here may be simply reduced to original entropic cosmological models associated with standard Hawking temperature and Bekenstein entropy, irrespective of the entropy definition.}

\subsection{Tsallis-Cirto Entropic Cosmology}

While Gibbs thermodynamics is based on the extensive and additive properties of the entropy $S$ of a composite thermodynamic system, Bekenstein entropy is instead nonextensive due to its area scaling and nonadditive. This means that it departs from the fundamental tenet of classical Gibbs thermodynamics if we consider black holes as $3+1$ dimensional objects. Therefore, black hole entropy and temperature definitions must be changed in order to fulfill the basic thermodynamic principles. 

In order to tackle this problem, Tsallis and Cirto \cite{Tsallis:2012js} start with the standard thermodynamic Legendre transformation, by which the Gibbs free energy $G$ for a general $d$ dimensional system reads as
\be
G=U-TS+pV-\mu N. \label{gibbs}
\ee
In the above equation, $U$, $S$, $V$, and $N$ stand for internal energy, entropy, volume, and particle number, respectively, while $T$, $p$, and $\mu$ denote temperature, pressure, and chemical potential.
The extensive variables $S$, $V$, and $N$ scale with $V=L^d$, where $L$ is the system's linear dimension. The intensive variables $T$, $p$, and $\mu$ scale with $L^\theta$. And finally the variables representing the energies, $G$ and $U$ scale with $L^\epsilon$. 
It follows from Eq.~(\ref{gibbs}) that
\be
\epsilon=\theta+d. \label{gibbs1}
\ee 
For a Schwarzschild black hole, $E\propto M \propto L$. In this situation, $\epsilon=1$, therefore we get from Eq.~(\ref{gibbs1}) that $\theta=1-d$. 

This shows that if we consider a black hole to be a two-dimensional object, the quantities $S_{BH}$ and $T_{BH}$ satisfy the Legendre structure, Eq.~(\ref{gibbs}). In fact,
the Bekenstein entropy will scale as $L^2$, and the temperature of the Schwarzschild black hole as $L^{-1}$, which is also the case for the Hawking temperature $T_{BH}$. 

But if we think of black holes as $3$-dimensional objects, based on the aforementioned Legendre structure, the definitions of both entropy and temperature must change. To overcome this issue, Tsallis and Cirto presented a new type of black hole entropy, which is described as follows:
\be
S_{TC}=k_B\left(\frac{S_{BH}}{k_B}\right)^\delta, 
 \label{S_Tsallis}
\ee 
with a new parameter $\delta=d/(d-1)>0$, and the corresponding composition rule given by
\begin{equation}   S_{TC,12}=\left[(S_{TC,1})^{1/\delta}+(S_{TC,2})^{1/\delta}\right]^\delta. \label{S_Tsallis_comp}
\end{equation}
By looking at the composition rule, Eq.~(\ref{S_Tsallis_comp}), it is clear that $S_{BH}$ is additive whereas $S_{TC}$ is nonadditive. Moreover, only for $\delta = 3/2$ $(d=3)$, $S_{TC}$ is proportional to the volume and, therefore, it is an extensive variable. Furthermore, in this case $T_{TC}$ must scale with $L^{-2}$. Employing Clausius' relation, Eq.~(\ref{eq:clausius}), and using Eq.~(\ref{S_Tsallis}) one can write the consistent Tsallis-Cirto temperature $T_{TC}$ as \cite{Cimdiker:2022ics}
\be
T_{TC}=\frac{T_{BH}}{\delta}\left(\frac{S_{BH}}{k_B}\right)^{1-\delta}. \label{T_Tsallis}
\ee 
which scales as $L^{-2}$ for $\delta=3/2$, i.e., $T_{TC} \propto M^{-2}$, for the case of a Schwarzschild 3D black hole. 
Using the previously defined nomenclature, Eqs.~(\ref{S_Tsallis})~-~(\ref{T_Tsallis}) correspond to the \textit{standard} scenario. One can easily check also that the entropic force in this case results to be
\be
F_{TC}=-T_{TC}\frac{dS_{TC}}{dL} = -\frac{c^4}{G}, \label{F_Tsallis}
\ee
and the entropic pressure
\be
p_{e}=\frac{F_{TC}}{A}=-\frac{c^2}{4\pi G}H^2. \label{p_Tsallis}
\ee
These results also show that the thermodynamically consistent Tsallis-Cirto entropic force models are identical to the entropic force models associated with Bekenstein entropy and Hawking temperature. 

In the \textit{horizon-scaled} scenario, performing the same checks and calculations, we get
\begin{align}
S^{\gamma}_{TC}&=k_B\left(\frac{S^{\gamma}_{BH}}{k_B}\right)^\delta, \label{S_Tsallis_gamma} \\
T^{\gamma}_{TC}&=\frac{T^{\gamma}_{BH}}{\delta}\left(\frac{S^{\gamma}_{BH}}{k_B}\right)^{1-\delta}\, , \label{T_Tsallis_gamma}
\end{align}
with the entropic force being
\be
F^{\gamma}_{TC}=-T^{\gamma}_{TC}\frac{dS^{\gamma}_{TC}}{dL} = -\frac{c^4}{G}\, , \label{F_Tsallis_gamma}
\ee
and the entropic pressure
\be
p^{\gamma}_{e}=\frac{F^{\gamma}_{TC}}{A}=-\gamma^2 \frac{c^2}{4\pi G}H^2. \label{p_Tsallis_gamma}
\ee
Finally, in the \textit{mass-scaled} scenario, we have that the consistent thermodynamical pair to be used is
\begin{align}
S^{M_\gamma}_{TC}&= S_{TC}\, , \label{S_Tsallis_Mgamma}\\
T^{M_\gamma}_{TC}&=\gamma\, T_{TC}\, , \label{T_Tsallis_Mgamma}
\end{align}
with the entropic force finally being
\be
F^{M_{\gamma}}_{TC}=-T^{M_\gamma}_{TC}\, \frac{dS^{M_\gamma}_{TC}}{dL} = -\gamma \frac{c^4}{G}\, , \label{F_Tsallis_Mgamma}
\ee
and the entropic pressure
\be
p^{M_{\gamma}}_{e}=\frac{F^{M_{\gamma}}_{TC}}{A}=-\gamma \frac{c^2}{4\pi G}H^2. \label{p_Tsallis_Mgamma}
\ee
Thus, finally, the entropic force and pressure are totally equivalent to same scenarios in the case of Bekenstein entropy and Hawking temperature.

In \cite{Komatsu:2013qia, Gohar:2020bod}, the authors model their generalized entropic force scenario defining $S_{TC}$ on the Hubble horizon for the particular value of $\delta=3/2$, which resulted in solving the issues of accelerating and decelerating universe and the problem of structure formation highlighted in \cite{Basilakos:2012ra, Basilakos:2014tha}. But in those studies they relate it to 
the Hawking temperature $T^{\gamma}_{BH}$: here we argue that \textit{Hawking temperature is not a good choice when employing Tsallis-Cirto entropy since it does not satisfy the Legendre structure, Eq.~(\ref{gibbs})} \cite{Cimdiker:2022ics,Nojiri:2021czz}. In fact, $S_{TC}$ with $\delta=3/2$ scales as $L^{3}$ and $T_{BH}$ scales as $L^{-1}$, while the proper corresponding temperature should scale as $L^{-2}$, which exactly happens with $T_{TC}$, Eq.~(\ref{T_Tsallis}).

\subsection{Tsallis-Zamora Entropic Cosmology}

Recently, Zamora and Tsallis \cite{Zamora:2022sya,Zamora:2022cqz} defined a new type of entropy and temperature for cosmological horizons, which is consistent from a thermodynamic point of view. The main motivation for their model are similar to the discussion in the previous section for the Tsallis-Cirto black hole entropy. The Tsallis-Zamora entropy $S_{TZ}$ and the corresponding temperature $T_{TZ}$ are given as
\begin{align}
    S_{TZ}&=\pi k_B\left(\frac{L}{l_p}\right)^d, \label{S_Zamora}\\
    T_{TZ}&=\frac{T_p}{\pi\,d}\left(\frac{L}{l_p}\right)^{1-d}\, , \label{T_Zamora}
\end{align}
where it is considered that the generalized entropy scales with the characteristic horizon with some arbitrary positive real power $d$. For $d=2$, $S_{TZ}$ and $T_{TZ}$ reduce to the standard Bekenstein entropy $S_{BH}$ and the Hawking temperature $T_{BH}$. 

The entropic force $F_{TZ}$ and entropic pressure $p_{e}$ for this case  become
\begin{align}
    F_{TZ}&=-\frac{c^4}{G}\, , \label{F_Zamora} \\
    p_e&=-\frac{c^2}{4\pi G}H^2\, . \label{p_Zamora} 
\end{align}
Thus also this entropic force model is formally equivalent to the Bekenstein entropic force models.

It is then straightforward to extend the Tsallis-Zamora model to the \textit{horizon-scaled} and \textit{mass-scaled} scenarios and to verify that, once again, they are equivalent to the Bekenstein-Hawking case at cosmological scales. In the former one, we have:
\begin{align}
    S^{\gamma}_{TZ}&=\frac{S_{TZ}}{\gamma^d},\label{S_Zamora_gamma}\\
    T^{\gamma}_{TZ}&= \frac{T_{TZ}}{\gamma^{1-d}}\, , \label{T_Zamora_gamma} 
\end{align}
leading to entropic force and pressure
\begin{align}
    F^{\gamma}_{TZ}&=-\frac{c^4}{G}\, , \label{F_Zamora_gamma} \\
    p^{\gamma}_e&=-\gamma^2 \frac{c^2}{4\pi G}H^2\, . \label{p_Zamora_gamma} 
\end{align}
In the latter one, instead, we have:
\begin{align}
    S^{M_\gamma}_{TZ}&= S_{TZ}, \label{S_Zamora_Mgamma}\\
    T^{M_{\gamma}}_{TZ}&=\gamma\, T_{TZ}\, , \label{T_Zamora_Mgamma}
\end{align}
with the corresponding entropic forse and pressure
\begin{align}
    F^{M_\gamma}_{TZ}&=-\gamma\,\frac{c^4}{G}\, , \label{F_Zamora_Mgamma} \\
    p^{M_\gamma}_e&=-\gamma\, \frac{c^2}{4\pi G}H^2\, . \label{p_Zamora_Mgamma} 
\end{align}
Since in its original formulation \cite{Zamora:2022cqz,Zamora:2022sya}, namely Eqs.~(\ref{S_Zamora})~-~(\ref{T_Zamora}), this model is equivalent to the original entropic force models, it also means that it is plagued by the same problems. Thus, in order to tackle this, an additional term depending on a smaller power $\Delta$, with $0<\Delta<d$, is added to the thermodynamically consistent generalized entropy $S_{TZ}$ in Eq.~(\ref{S_Zamora}), which is given by \cite{Zamora:2022cqz}
\begin{equation}
    S_{TZ,mod}=\pi k_B\left(\frac{L}{lp} \right)^d+\frac{k_B}{\Delta}\left[\left(\frac{L}{l_p}\right)^\Delta-1\right]\, . \label{S_Zamora_mod}
\end{equation}
But then, we assert here that the considerations made by \cite{Zamora:2022cqz} are not strictly valid. Indeed, they assume that $S_{TZ,mod}$ scales as $L^d$ and therefore \cite{Zamora:2022cqz} suggest that $T_{TZ}$ is used to define the entropic force $F_{TZ,mod}$, which reads as 
\begin{equation}
    F_{TZ,mod}=-d\frac{c^4}{G}(1+\alpha H^{d-\Delta}),
\end{equation}
where $\alpha=(l_p/c)^{d-\Delta}/d$. But this is not correct and it might be considered just as an approximation valid in the case $\Delta\ll 1$ (which should be, anyway, verified by testing with data, or at least quantitavely estimated by robust theoretical considerations). Moreover considering that $T_{TZ}$ scales as $L^{1-d}$, it is clear that $S_{TZ,mod}$ and $T_{TZ}$ do not satisfy the Legendre structure.

Actually, the correct corresponding temperature to be used should be
\begin{equation}
    T_{TZ,mod}=T_p \left(\frac{L}{l_p}\right)^{1-d} \left[\left(\frac{L}{l_p} \right)^{\Delta-d}+\pi\,d\right]^{-1}\, , \label{T_Zamora_mod}
\end{equation}
from which the calculations of the entropic force and pressure finally give
\begin{align}
    F_{TZ,mod}&=-\frac{c^4}{G}\, , \label{F_Zamora_mod} \\
    p_{e,TZ,mod}&=-\frac{c^2}{4\pi G}H^2\, . \label{p_Zamora_mod} 
\end{align}
Once again, we end up with an entropic force model which is formally equivalent to the Bekenstein entropic force models.

Finally, we can verify that also in the \textit{horizon-scaled} and \textit{mass-scaled} scenarios the equivalence with the original entropic models is saved. Indeed, in the former case we have
\begin{align}
S^{\gamma}_{TZ,mod}&=\pi k_B\left(\frac{L}{\gamma\,l_p} \right)^d+\frac{k_B}{\Delta}\left[\left(\frac{L}{\gamma\,l_p}\right)^\Delta-1\right], \label{S_Zamora_mod_gamma}\\
T^{\gamma}_{TZ,mod}&=T_p \left(\frac{L}{\gamma\,l_p}\right)^{1-d} \left[\left(\frac{L}{\gamma\,l_p} \right)^{\Delta-d}+\pi\,d\right]^{-1}, \label{T_Zamora_mod_gamma}
\end{align}
with entropic force and pressure being
\begin{align}
    F^{\gamma}_{TZ,mod}&=-\frac{c^4}{G}\, , \label{F_Zamora_mod_gamma} \\
    p^{\gamma}_{e,TZ,mod}&=-\gamma^2\frac{c^2}{4\pi G}H^2\, . \label{p_Zamora_mod_gamma} 
\end{align}
While in the former one we have
\begin{align}
S^{M_\gamma}_{TZ,mod}&= S_{TZ,mod}, \label{S_Zamora_mod_Mgamma}\\
T^{M_\gamma}_{TZ,mod}&=\gamma T_{TZ,mod}, \label{T_Zamora_mod_Mgamma}
\end{align}
with entropic force and pressure
\begin{align}
    F^{M_\gamma}_{TZ,mod}&=-\gamma\, \frac{c^4}{G}\, , \label{F_Zamora_mod_Mgamma} \\
    p^{M_\gamma}_{e,TZ,mod}&=-\gamma\, \frac{c^2}{4\pi G}H^2\, . \label{p_Zamora_mod_Mgamma} 
\end{align}
Again, we have the entropic force and the pressure similar to the standard case of Bekenstein and Hawking.

\subsection{R\'enyi Entropic Cosmology}

R\'enyi defined a generalization of Shannon entropy that depends on a parameter $q$ in 1960 \cite{reny1}.
The definition of the R\'enyi entropy of order $q$ for a probability distribution $p$ on a finite set is given by
\be
S_R= k_B \frac{\ln \sum_i p^q(i)}{1-q}. \label{S_R}
\ee
such that, in the limit $q \to 1$, Eq.~(\ref{S_R}) reduces to Shannon's entropy
\be
S_G=-k_B\sum_ip(i)\ln p(i). \label{SG}
\ee
In this way, Tsallis entropy $S_q$ is a generalization of Shannon's entropy $S_G$, which can be written as
\be
S_q=-k_B\sum_{i}[p(i)]^q\ln_qp(i), \label{S_T}.
\ee 
The parameter $q$ specifies the degree of nonextensivity, and we consider it positive to assure the concavity of $S_q$, where $p(i)$ is the probability distribution specified on a collection of microstates $\Omega$. The definition of the q-logarithmic function $\ln_q p$ is given by
\be
\ln_qp=\frac{p^{1-q}-1}{1-q},
\ee 
such that, again, in the limit $q \to 1$, Eq.~(\ref{S_T}) reduces to Shannon's entropy.
It should be noted that Shannon's entropy, Eq.~ (\ref{SG}), satisfies the additive composition rule, whereas Tsallis entropy, Eq.~(\ref{S_T}), satisfies a nonadditive composition rule. However, using the ``formal logarithm'' technique, as described in \cite{Biro:2013cra}, one can formulate R\'enyi entropy $S_R$ in terms of $S_q$ such that the Eq.~(\ref{S_R}) becomes
\be
S_R=\frac{k_B}{1-q}\ln{\left(1+\frac{1-q}{k_B}S_q\right)},
\label{ERenyi}
\ee
In \cite{Cimdiker:2022ics}, it is shown that, for every nonextensive entropy, one can define an equilibrium temperature by utilizing the equilibrium condition by maximizing the nonextensive entropy and that there is always an additive equilibrium entropy associated with this equilibrium temperature. In this regard, the equilibrium temperature in the Tsallis nonextensive setup is given by \cite{Cimdiker:2022ics}
\be
T_R=\left(1+\frac{1-q}{k_B} S_{q}\right)\frac{1}{k_B\beta}, \label{T_R}
\ee
which is exactly the R\'enyi temperature corresponding to $S_R$. This means that the $S_R$ is the equilibrium entropy in the Tsallis nonextensive set up. Note that in Eq.~(\ref{T_R}), $k_B\beta =\frac{\partial S_q}{\partial U}$, where $U$ is the internal energy of the Tsallis nonextensive system.

Since Bekenstein entropy $S_{BH}$ is considered to be nonextensive, therefore, to employ R\'enyi entropy $S_R$, Eq.~(\ref{S_R}), the main assumption we consider here is that $S_{BH}=S_q$ in Eq.~(\ref{ERenyi}) and, therefore, for the case of black hole or cosmological horizons, it reads as
\begin{equation}\label{S_Renyi}
  S_{R}=\frac{k_B}{\lambda}\ln \left(1+\frac{\lambda}{k_B} S_{BH}\right),
\end{equation}
where we introduce $\lambda=1-q$. Using Clausius's relation, one can write the corresponding temperature $T_R$ as
\begin{equation}\label{T_Renyi}
T_R=\left(1+\frac{\lambda}{k_B} S_{BH}\right)T_{BH}.
\end{equation}
The entropic force and pressure, $F_R$ and $p_R$, become again
\begin{align}
  F_{R}&=-T_{R}\frac{dS_{R}}{dL}= - \frac{c^4}{G}, \label{F_Renyi} \\
  p_{R}&=-\frac{c^2}{4\pi G}H^2 \label{p_Renyi}
\end{align}
which means that the R\'enyi entropic force model with the consistent thermodynamic quantities gives exactly the same model than in the case of Bekenstein entropy and the Hawking temperature.

As expected, the same results can be found in the \textit{horizon-scaled} and \textit{mass-scaled} cases. In the first case we have
\begin{align}
  S^{\gamma}_{R}&=\frac{k_B}{\lambda}\ln \left(1+\frac{\lambda}{k_B} S^{\gamma}_{BH}\right), \label{S_Renyi_gamma} \\
  T^{\gamma}_R&=\left(1+\frac{\lambda}{k_B} S^{\gamma}_{BH}\right)T^{\gamma}_{BH}, \label{T_Renyi_gamma} 
\end{align}
leading to entropic force and pressure
\begin{align}
  F^{\gamma}_{R}&=-T^{\gamma}_{R}\frac{dS^{\gamma}_{R}}{dL}= - \frac{c^4}{G}, \label{F_Renyi_gamma} \\
  p^{\gamma}_{R}&=-\gamma^2 \frac{c^2}{4\pi G}H^2. \label{p_Renyi_gamma}
\end{align}
And in the second one we get
\begin{align}
  S^{M_\gamma}_{R}&=S_{R}, \label{S_Renyi_Mgamma} \\
    T^{M\gamma}_R&=\left(1+\frac{\lambda}{k_B} S^{\gamma}_{BH}\right)T^{M_\gamma}_{BH}, \label{T_Renyi_Mgamma} 
\end{align}
leading to
\begin{align}
  F^{M_\gamma}_{R}&=-T^{M_\gamma}_{R}\frac{dS^{M_\gamma}_{R}}{dL}= - \gamma \frac{c^4}{G}, \label{F_Renyi_Mgamma} \\
  p^{M_\gamma}_{R}&=-\gamma \frac{c^2}{4\pi G}H^2. \label{p_Renyi_Mgamma}
\end{align}
This again shows that the R\'enyi entropic force models are similar to the standard Bekenstein-Hawking entropic force models.

\section{Discussion and Conclusions}
\label{sec:discussion}

{\it ``The present paper lays no claim to deep originality. Its main purpose is to give a systematic treatment''} \cite{Chevalley:1948zz} of many theoretical and methodological inconsistencies which seem to plague the recent literature of entropic cosmology and other thermodynamic approaches to cosmology, and which in our opinion it fundamental to solve because they are directly related to the foundations of thermodynamic relations applied to black hole and cosmological horizons. 

By using ``standard'' Bekenstein entropy $S_{BH}$ and Hawking temperature $T_{BH}$, we began by highlighting the two main steps to be followed in order to consistently construct an entropic model. 

\texttt{Remark nr.~1.} The first crucial fundamental need is that temperature and entropy must obey the Clausius relation $dE=TdS$. \textit{If any newly defined pair of temperature and entropy does not satisfy Clausius relation, the model is thermodinamically inconsistent and flawed since the beginning.}

\texttt{Remark nr.~2.} \textit{Then, one must propose a relation to connect energy, mass and the cosmological horizon.} In the standard case, one proceeds by analogy from the fact the a Schwarzschild black hole with mass $M$ has an event horizon $L= 2GM/c^2$, i.e. the mass $M$ linearly scales with the horizon $L$, which is pretty evident in the context of black holes, but it is just an assumption when working with cosmological horizon. 

Joining the Clausius relation and the $M-L$ black holes relations, one then can derive that the standard Bekenstein entropy, $S_{BH}$, and Hawking temperature, $T_{BH}$, are a thermodinamically consistent pair on which to build up an entropic cosmology (of course, one should then to check if they can really mimic and substitute dark energy). This also implies that we can derive $T_{BH}$ or $S_{BH}$, respectively, when either is provided, if we rely on both the Clausius relation and the black holes $M-L$ relation. 

\texttt{Remark nr.~3.} We have provided a physical meaning to the Hawking temperature parameter $\gamma$ in the context of entropic cosmology, which is overlooked and not correctly addressed in entropic cosmology works. We have demonstrated that \textit{the only way to introduce it consistently in the construction of an entropic model based on $S_{BH}$ and $T^{\gamma}_{BH}$, is to have $\gamma$ controlling the scaling between $M$ and $L$}, i.e., $M = \gamma c^2L/G$, and not, as in the original intention, ``quantifying our ignorance of the cosmological horizon''. In such a way the parameter $\gamma$ seems to be also more naturally connected to more fundamental issues. Indeed, in this case it provides a way to parameterize the slope of the $M-L$ relation, which would be otherwise fixed to $c^2/G$. 

On the other hand, if one would still retain such original meaning for $\gamma$, then everything, i.e Bekenstein entropy, Hawking temperature, and the mass-horizon relation, should be properly rescaled.

\texttt{Remark nr.~4.} In recent years, different nonextensive entropy proposals have been applied to gravity and cosmological applications. In this paper, we have investigated nonextensive entropic cosmology within the nonextensive setup, more specifically looking at Tsallis-Cirto entropy, R\'enyi entropy, and Tsallis-Zamora entropy. \textit{We have demonstrated that, with consistent thermodynamic quantities defined on the Hubble horizon, satisfying Clausius relation, and employing a linear $M-L$ relation, all the investigated nonextensive entropic force models are identical to the original entropic force models that are derived from the standard Bekenstein entropy and the Hawking temperature.} And very likely this can be extended to any other newly proposed functional entropic form.

Thus, any claim stating that they might provide different solutions on the cosmological level is simply wrong. Actually, the main mistake in literature has been to couple these new nonextensive entropies with the standard Hawking temperature, but this has an intrinsic fundamental flaw, being the corresponding entropy-temperature pairs not consistent with Clausius relation.

\texttt{Remark nr. 5} With the holographic principle taken into account, the entropy and temperature defined on the Hubble horizon give the entropic force on the horizon, which is thought to be the cause of the universe's acceleration in the context of entropic cosmology. Intriguingly, we have demonstrated in this paper that the entropic force acting on the universe's boundary for the cases of Bekenstein, Tsallis-Cirto, Tsallis-Zamora, and R\'enyi entropy gives $\gamma c^4/G$ while using the corresponding consistent temperature. \textit{This demonstrates that no matter what definitions of entropies and temperatures are used in the definition of entropic force, the entropic force will remain the same if we use the consistent thermodynamic quantities on the horizons, and its form is mostly dictated by the adopted mass-to-horizon relation.} This is the main reason for the Tsallis-Cirto, Tsallis-Zamora, and R\'enyi entropic force models are identical to the Bekenstein entropic force model when using consistent quantities. 

\texttt{Remark nr.~6.} Given the previous consideration,  \textit{there is no way to describe cosmological data and the acceleration of our Universe in entropic cosmologies approach if we rely only on proposing new definitions for the entropy}. 

In fact, it has been already demonstrated in the literature that an entropic force term $\propto H^2$, as it comes out from standard Bekenstein entropy, is unable to describe both the cosmological background (thus, data like those from Type Ia Supernovae, or from Baryon Acoustic Oscillations) and the growth of cosmological perturbations (thus, the Cosmic Microwave Background, or the Redshift Space Distortion data from matter distribution on large scales). 

A possible way-out could be provided by the proper introduction of the parameter $\gamma$. What is mostly crucial for us to address here, is that if any further investigation  should be address in the field of entropic cosmology, with the goal of providing a fully reliable natural explanation for dark energy, new paths must be explored and new ideas, involving questioning at the very fundamental level, must be proposed. In this regard, we have investigated all these issues in \cite{HG}.

\bibliographystyle{apsrev4-1}
\bibliography{ref1}

\end{document}